\documentclass[sigconf]{aamas} 

\usepackage{balance} 

\setcopyright{ifaamas}
\acmConference[AAMAS '24]{Proc.\@ of the 23rd International Conference
on Autonomous Agents and Multiagent Systems (AAMAS 2024)}{May 6 -- 10, 2024}
{Auckland, New Zealand}{N.~Alechina, V.~Dignum, M.~Dastani, J.S.~Sichman (eds.)}
\copyrightyear{2024}
\acmYear{2024}
\acmDOI{}
\acmPrice{}
\acmISBN{}

\acmSubmissionID{253}

\title{Towards Generalizability of Multi-Agent Reinforcement Learning in Graphs with Recurrent Message Passing}

\author{Jannis Weil}
\affiliation{
  \institution{Technical University of Darmstadt}
  \city{Darmstadt}
  \country{Germany}}
\email{jannis.weil@tu-darmstadt.de}

\author{Zhenghua Bao}
\affiliation{
  \institution{Technical University of Darmstadt}
  \city{Darmstadt}
  \country{Germany}}
  \email{zhenghua.bao@stud.tu-darmstadt.de}

\author{Osama Abboud}
\affiliation{
  \institution{Huawei Technologies}
  \city{Munich}
  \country{Germany}}
  \email{osama.abboud@huawei.com}
  
\author{Tobias Meuser}
\affiliation{
\institution{Technical University of Darmstadt}
\city{Darmstadt}
\country{Germany}
}
\email{tobias.meuser@tu-darmstadt.de}

\begin{abstract}
Graph-based environments pose unique challenges to multi-agent reinforcement learning.
In decentralized approaches, agents operate within a given graph and make decisions based on partial or outdated observations.
The size of the observed neighborhood limits the generalizability to different graphs and affects the reactivity of agents, the quality of the selected actions, and the communication overhead.
This work focuses on generalizability and resolves the trade-off in observed neighborhood size with a continuous information flow in the whole graph.
We propose a recurrent message-passing model that iterates with the environment's steps and allows nodes to create a global representation of the graph by exchanging messages with their neighbors.
Agents receive the resulting learned graph observations based on their location in the graph.
Our approach can be used in a decentralized manner at runtime and in combination with a reinforcement learning algorithm of choice.
We evaluate our method across 1000 diverse graphs in the context of routing in communication networks and find that it enables agents to generalize and adapt to changes in the graph.

\end{abstract}

\keywords{Multi-Agent Reinforcement Learning; Graph Neural Networks; Communication Networks}

\newcommand{\BibTeX}{\rm B\kern-.05em{\sc i\kern-.025em b}\kern-.08em\TeX}
\usepackage{hyperref}

\usepackage{tikz}

\usepackage{adjustbox}
\newcommand{\marktext}[2]{\adjustbox{bgcolor=#1}{\strut #2}}

\usepackage{caption}
\usepackage{subcaption}

\usepackage{booktabs}
\usepackage{multirow}

\usepackage{arydshln} 
\usepackage{todonotes}

\usepackage[nolist]{acronym}
\usepackage{amsmath}
\usepackage{amsfonts}
\usepackage{mathtools}

\usepackage{algorithm}
\usepackage[noend]{algpseudocode}
\makeatletter
\let\OldStatex\Statex
\renewcommand{\Statex}[1][3]{%
  \setlength\@tempdima{\algorithmicindent}%
  \OldStatex\hskip\dimexpr#1\@tempdima\relax}
\makeatother

\makeatletter
\providetoggle{ALG@Ruler@empty}
\newcounter{ALG@Ruler@Id}
\apptocmd{\ALG@beginblock}{%
    \stepcounter{ALG@Ruler@Id}%
    \def\ALG@Ruler{ALG@Ruler@BeginPoint-\arabic{ALG@Ruler@Id}}%
    \expandafter\edef\csname ALG@Ruler@AtLevel@\theALG@nested\endcsname{\ALG@Ruler}%
    \tikz[remember picture, overlay] \coordinate (\ALG@Ruler);%
    \toggletrue{ALG@Ruler@empty}%
}{}{}
\pretocmd{\ALG@endblock}{%
    \iftoggle{ALG@Ruler@empty}{}{%
        \def\ALG@Ruler{\csname ALG@Ruler@AtLevel@\theALG@nested\endcsname}%
        \unskip\tikz[remember picture, overlay] \draw[black!25]
            ([xshift=0.5em, yshift=-0.5ex]\ALG@Ruler) |- ([xshift=1em, yshift=-1.0ex]\ALG@Ruler|-0,0);%
    }%
    \togglefalse{ALG@Ruler@empty}%
}{}{}
\makeatother

\newcommand{\algemph}[1]{\marktext{lightgray!40}{\strut #1}}
\newcommand{\algemphmath}[1]{\marktext{lightgray!40}{\strut $\displaystyle #1$}}

\DeclareMathOperator{\mathdef}{ \dot{=} }
\DeclareMathOperator{\mathconcat}{\mathbin\Vert}

\DeclareMathOperator{\mathnat}{ \mathbb{N} }

\usepackage{xspace}

\begin{acronym}[decpomdp]
\acro{decpomdp}[Dec-POMDP]{Decentralised Partially Observable Markov Decision Process}
\acro{qoe}[QoE]{Quality of Experience}
\acro{sdn}[SDN]{Software-defined Networking}
\acro{marl}[MARL]{Multi-Agent Reinforcement Learning}
\end{acronym}

\makeatletter
\gdef\@copyrightpermission{
	\begin{minipage}{0.3\columnwidth}
		\href{https://creativecommons.org/licenses/by/4.0/}{\includegraphics[width=0.90\textwidth]{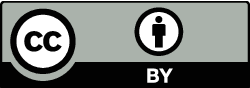}}
	\end{minipage}\hfill
	\begin{minipage}{0.7\columnwidth}
		\href{https://creativecommons.org/licenses/by/4.0/}{This work is licensed under a Creative Commons Attribution International 4.0 License.}
	\end{minipage}
	\vspace{5pt}
}
\makeatother

\begin{document}


\pagestyle{fancy}
\fancyhead{}


\maketitle 


\section{Introduction}

The capability of an adaptive system depends on the quality of its input.
Ideally, it has access to the state and makes fully informed decisions at all times.
Research in multi-agent reinforcement learning ranges from centralized to decentralized approaches~\cite{li2022MARLinFutureInternet}.
Our focus lies on decentralized systems in graph-based environments.
While agents may leverage the complete state during training, they have limited access to local information during execution.
In decentralized approaches, agents can directly react to local changes.
However, having only local information may lead to suboptimal decisions.
A common way to counteract this is to expand the observations of each agent by  information from their direct neighborhood~\cite{milena20CognitiveCache}.
Including more nodes in the observed neighborhood improves decision making~\cite{brandherm2022bigMEC} but increases the communication overhead.

Recent works show that graph neural networks \cite{scarselli2009gnn} and neural message passing \cite{gilmer2017neuralMessagePassing} are well suited for applications in graph-based environments, especially because they can generalize to unseen graphs~\cite{rusek2020routeNetGNNModelling}.
However, the approaches often assume a centralized view \cite{almasan2022gnnRouting}, explicit coordination across all agents \cite{bernardez2021MLTrafficOptim}, or the availability of labeled data in order to apply supervised learning~\cite{geyer2018distributedRouting}.

We aim to resolve the trade-off in limited observed neighborhoods in graph-based environments and propose a recurrent approach where nodes exchange local information via message passing to improve their understanding of the global state.
They refine their local states over the environment's steps, allowing information to iteratively travel through the whole graph.
Based on these node states, agents receive location-dependent graph observations.

Our approach provides a novel foundation for learning communication systems in multi-agent reinforcement learning and is jointly trained in an end-to-end fashion.
Our contributions are:
\begin{itemize}
    \item We propose to \emph{decouple} learning graph-specific representations and control by separating node and agent observations. 
    \item To the best of our knowledge, we are the first to address the problem of limited observed neighborhoods in graph-based environments with recurrent graph neural networks.
    \item We show that the learned graph observations enable generalization over 1000 diverse graphs in a routing environment, achieving similar throughput as agents that specialize on single graphs when combined with action masking.
    \item We show that our approach enables agents to adapt to a change in the graph on the fly without retraining.
\end{itemize}

Our code is available at \href{https://github.com/jw3il/graph-marl}{https://github.com/jw3il/graph-marl}.
The remainder of this paper is structured as follows.
We begin with the problem statement in Sec.~\ref{sec:problem-statement} and then introduce our approach in Sec.~\ref{sec:approach}.
We describe our evaluation setup in Sec.~\ref{sec:experiment-setup}, the results are presented and discussed in Sec.~\ref{sec:experiment-results}.
The following Sec.~\ref{sec:related-work} provides an overview of
related work and Sec.~\ref{sec:conclusion} concludes the paper.

\section{Reinforcement Learning in Graphs}
\label{sec:problem-statement}

\looseness=-1
Reinforcement learning (RL) in graph-based environments has gained much popularity in many application domains \cite{nie23graphRL}, including communication networks~\cite{li2022MARLinFutureInternet}.
We consider multi-agent environments that build upon a graph $G \doteq (V,\, E) \in \mathcal{G}$, representing a communication network of nodes $V$ connected via undirected edges $E \subseteq V \times V$.
We assume that the agents $I$ are part of a partially observable stochastic game~\cite{hansen04DynamicProgrammingPOSG} that requires them to consider the graph, i.e.\ state $s_t\in S$ at step $t$ contains $G$ and may augment it with state information that characterizes the network.
Examples include edge delays and compute resources of nodes.
Each agent $i \in I$ receives partial observations $o^i_t\sim O^i(\cdot \vert s_t)$ and selects an action $a^i_t\in A$ using its policy $a^i_t \sim \pi^i(\cdot \vert o^i_t)$.
The agent's goal is to maximize its expected discounted return $\mathbb{E}\bigl[ \sum_{t'=t}^T \gamma^{t' - t} R^i_{t'} \bigr]$ with discount factor $\gamma \in [0, 1]$ and time horizon $T$, where the individual rewards $R^i_t = R^i(s_t, a_t)$ are based on the joint action of all agents $a_t \mathdef{} (a^i_t)_{i \in I}$.

How agents observe the graph and the network's state is usually not discussed in depth by related works.
A centralized view allows for the best decision making but does not scale to bigger graphs.
A decentralized view trades off reactivity and amount of available information with the size of the observed neighborhood.
Authors usually decide for one of these views and their approaches are therefore, by design, limited to certain problems or graph structures.

Consider a decentralized approach where an agent is located on a node and observes its $n$-hop neighborhood.
Depending on the environment, the assumption about the observable neighborhood is critical with respect to generalizability.
Let's imagine the agent is supposed to find the shortest path to some destination node in the graph that is $n+1$ hops away.
How can it identify the shortest path to a destination node that's not included in its observation?
There are two straight-forward solutions:
\begin{enumerate}
    \item Specialization on a single graph. Agents explore the whole graph during training. If the graph is fixed and nodes are uniquely identifiable based on the agent's observation, agents can specialize and find the optimal path to any node.
    \item Expansion of the observation space to include the missing information, e.g. to $(n+1)$-hop neighborhoods.
\end{enumerate}

With specialization, the learned solution will not generalize to other graphs.
If the target graph does not change, this would be a sufficient but highly inflexible solution.
As real networks have diverse underlying graphs and are usually dynamic, many researchers consider online training on the target graph.
However, as reinforcement learning requires agents to explore and make suboptimal decisions, online training from scratch might be unacceptable in practice.
In contrast, expanding the observation space allows agents to generalize.
However, the observation range to consider greatly depends on the concrete problem.
For example, for routing, global knowledge is necessary if any node could be the destination.
But then the approach is not decentralized anymore and will not scale.

Our idea is to address this issue by expanding the observation space with learned graph observations that leverage recurrent message passing.
Agents are still reactive and don't have to gather information about the whole graph before making a decision.
While initial decisions may be suboptimal, the quality of the learned graph observations should increase over time.
Ideally, they converge to a global view and allow agents to make optimal decisions.

\section{Learned Graph Observations}
\label{sec:approach}

\begin{figure}[!h]
    \centering
    \includegraphics[width=\linewidth]{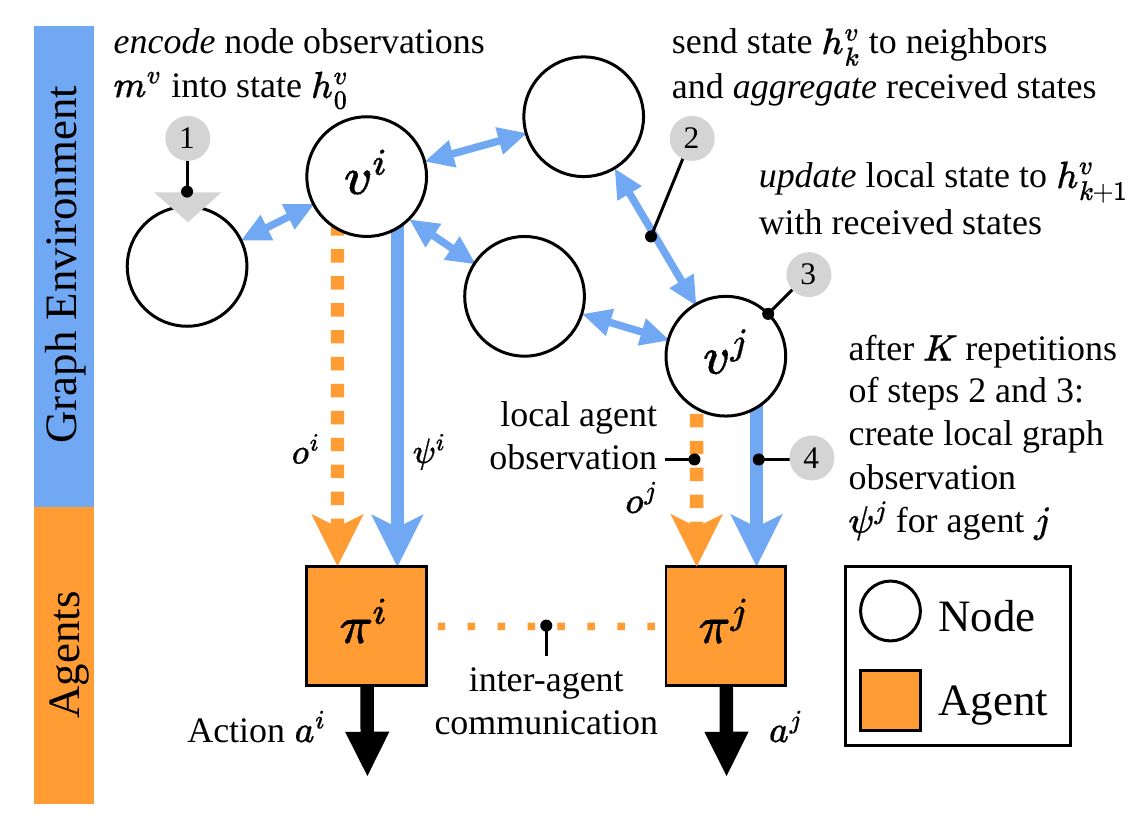}
    \vspace{-0.4cm}
    \caption{Our graph observation mechanism iteratively distributes node states via message passing. Agents in the graph receive local graph observations for decision making.}
    \Description{The figure shows a graph with five nodes at the top and two agents at the bottom. Each agent is assigned to one node in the graph. In the graph, the steps of recurrent message passing are annotated. Step 1 with encode points to a node, step 2 with aggregate points at an edge between two nodes, step 3 with update points at a node and step 4 with the readout function points at the connection from a node to an agent. Agents receive graph observations from nodes and local agent observations from the environment and may communicate via inter-agent communication. Agents output individual actions.}
    \vspace{-0.25cm}
    \label{fig:monitoring-concept}
\end{figure}

We consider environments that are based on a graph and propose to \emph{decouple} both components, i.e.~agents do not have to keep track of the whole graph state and can build upon a lower-level mechanism that aggregates graph information.
The information flow is illustrated in Fig.~\ref{fig:monitoring-concept} and will be explained in the following subsections.

\subsection{Recurrent Message Passing}
\label{sec:message-passing}

The core idea of our graph observations is leveraging the message-passing framework of graph neural networks \cite{hamilton2020GraphRepresentationLearning} to distribute local information in the network.
Related approaches are usually used in a centralized manner with a global view \cite{bhavanasi2023gcnRouting}. 
However, recurrent aggregation functions \cite{seo18gconvlstm, li16gatedGraphSequenceModels} spread multiple message passing iterations over time and enable decentralization \cite{geyer2018distributedRouting}.
We introduce a second recurrent loop back to the input of the graph neural network and label this approach \emph{recurrent message passing}.

We assume that each \textit{node} $v\in V$ receives local node observations  $m^v \sim M^v(\cdot \mid s)$ based on an unknown system state $s \in S$.
Instead of directly using this as an input of a graph neural network, each node $v\in V$ embeds its local observation into its current node state $h^v$ with an arbitrary differentiable function $\textit{encode}$:
\begin{equation}
    \label{eq:message-passing-encoding}
    h^v_0 \doteq \textit{encode}(h^v, m^v).
\end{equation}

The node state $h^v$ is initialized to zero in the first step and will serve as a recurrent loop between subsequent environment steps.
This allows nodes to consider previously aggregated information when processing observations, similar to auto-regressive graph models~\cite{rempe2022drivingScenarios} that use predictions of previous steps as their input.

At iteration $k \geq 0$, each node sends their state $h^v_k$ to all direct neighbors $w \in N(v) \doteq \{w \mid (v, w) \in E\}$.
Each neighbor generates a new state by first aggregating incoming node states and then updating its state using arbitrary differentiable functions $\textit{aggregate}^k$ and $\textit{update}^k$.
One message passing iteration is defined as:
\begin{align}
    \label{eq:message-passing-aggregate}
    M^v_k &\doteq \textit{aggregate}^k((h^w_k)_{w \in N(v)})\\
    \label{eq:message-passing-update}
    h^v_{k+1} &\doteq \textit{update}^k(h^v_k, M^v_k).
\end{align}

Alg.~\ref{alg:approach} shows pseudo code for recurrent message passing that is executed by all nodes $v \in V$.
There are $K \in \mathnat{}$ iterations between steps in the environment, i.e.\ equations (\ref{eq:message-passing-aggregate}) and (\ref{eq:message-passing-update}) are repeatedly applied until $k + 1 = K$.
The final aggregate $h^v_K$ and all intermediate node states $H^v$ received and calculated by $v$ can then be used by agents, as we will detail later. 
In the next environment step, we set $h^v = h^v_K$ and repeat the algorithm.
The number of iterations can be adjusted to fit the requirements of the learning task.
An increased number of iterations per step causes information to traverse the network faster but also increases the communication overhead.

\algrenewcommand\algorithmicindent{1.0em}%
\algrenewcommand\algorithmicrequire{\textbf{Input:}}
\algrenewcommand\algorithmicensure{\textbf{Output:}}
\begin{algorithm}[!t]
\caption{Distributed Node State Update}\label{alg:approach}
\begin{algorithmic}[1]
\Require Node $v$ with direct neighbors $N(v)$, state $s$, previous node state $h^v$, node observation $m^v$
\Ensure Updated node state $h^v_K$ and intermediate states $H^v$

\State $h^v_0 \leftarrow \textit{encode}(h^v, m^v)$
\Comment{Encode node observation}

\For{$k \leftarrow 0 \text{ to } K-1$}
\Comment{Update with message passing}
    \State Send $h^v_k$ to all neighbors $w \in N(v)$
    \State Receive $h^w_k$ from all neighbors $w \in N(v)$
    \State $M^v_k \leftarrow \textit{aggregate}^k((h^w_k)_{w \in N(v)})$
    \State $h^v_{k+1} \leftarrow \textit{update}^k(h^v_k, M^v_k)$
\EndFor
\State $H^v \leftarrow (h^v_k)_k \mathconcat{} (h^w_k)_{w\in N (v), k}$ \Comment{Get all intermediate states}
\State \textbf{return} $h^v_K, H^v$ 
\end{algorithmic}
\end{algorithm}

This node state update is performed at each step in the environment.
We assume that each agent $i\in I$ is assigned to exactly one node $v^i \in V$ at each step.
In addition to its observation in the environment, agent $i \in I$ then receives a local graph observation $\psi^i$ of the node it is assigned to based on the information $H^{v^i}$ this node received in this step via a differentiable readout function $\Psi$:
\begin{equation}
    \label{eq:message-passing-agent-obs}
    \psi^i \doteq \Psi(H^{v^i}).
\end{equation}
In the simplest case, $\Psi$ could be the identity function of the latest node state $\Psi(H^{v^i}) \doteq h^{v^i}_K$ in each iteration, i.e.\ agents receive the current state of the node to which they are assigned.
Access to intermediate node states is necessary to allow for skip connections and aggregation mechanisms like jumping knowledge networks~\cite{xu2018jumpingKnowledge}.

\subsection{Model Architecture}
\label{sec:design-architecture}
\begin{figure}[!t]
    \centering
    \includegraphics[width=\linewidth]{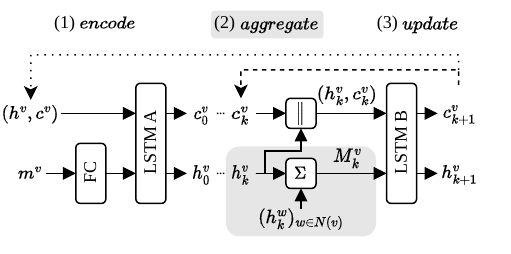}
     \caption{Our recurrent message passing model leverages LSTM cells to encode the node observation and update the node state. The hidden states of neighbor nodes are aggregated via summation, cell states remain local to each node.}
     \Description{The figure shows the architecture of our message passing model. The three stages are assigned to individual blocks of the model. Encode is assigned to a fully-connected network and an LSTM A, aggregate is assigned to the summation of the neighbors' hidden states and update is assigned to an LSTM B.}
    \label{fig:recurrent-message-passing}
\end{figure}

Based on the design from the previous section, we propose a simple recurrent message passing architecture.
The $\textit{encode}$ function is represented by a fully connected network to embed the node observation and an LSTM~\cite{schmidhuber1997lstm} to update the previous node state with the new embedding.
The $\textit{aggregate}$ function is the sum of all neighbors' hidden states, but could be any graph convolution from related work.
Finally, $\textit{update}$ is modeled by another LSTM.
We share parameters for all iterations, i.e. $\forall k.\: \textit{update}^k = \textit{update}$.
We provide an overview of the architecture with Fig.~\ref{fig:recurrent-message-passing}. 
An LSTM cell takes an input tensor and a pair of hidden and cell state tensors $(h, c)$ and yields new hidden and cell states.
In our architecture, the hidden state is exchanged with neighbor nodes during aggregation, the cell state remains local to the node.
The inner loop from update to aggregate depicts iterations within a step, the outer loop represents the forwarding of states between environment steps.
The states of the LSTM modules are not separated. 
Instead, a single pair of hidden and cell states is passed on between the modules.

\subsection{Integration in Deep RL}
Graph observations are compatible with all deep reinforcement learning approaches that allow for backpropagation through the policy's input, i.e.\ the observation space.
To the best of our knowledge, most algorithms do not have any limitations in that regard, as the policy is usually based on a differentiable function.

In order to integrate our method with deep RL algorithms, the node state update has to be performed at each environment step during inference and training, resulting in an expanded observation space for the agents.
The integration at inference time can be done with a simple environment wrapper.
Depending on the considered algorithm, the integration into the training loop will require additional effort.
For example, algorithms based on Q-learning bootstrap the value target using the observations of future states.
In order to compute the corresponding graph observations, we therefore have to compute or sample node states for these future steps.
Additionally, as node states are updated over multiple environment steps, we have to unroll the node state update over a sequence of steps and apply backpropagation through time in order to learn stable update functions.
This is already included in algorithms that consider stateful agents with recurrent models~\cite{kapturowski19R2D2}, but will require adjustments for other reinforcement learning algorithms.

\subsection{Exemplary Integration in Deep Q-Learning}
\label{sec:integration-dqn}
In this section, we exemplary describe how to integrate our method into independent DQN~\cite{mnih2015DQN} with parameter sharing across agents.
Our approach is summarized in Alg.~\ref{alg:psuedo-dqn}, noteworthy changes to the original algorithm are highlighted in light gray.
The main difference lies in the introduction of node states $h_t$ that are updated in parallel to the environment steps based on the node observations $m_t$ and the previous node states.
For notational simplicity, we denote recurrent message passing combined with the readout function $\Psi$ as a differentiable function $U(h_t, m_t, s_t; \theta_U)$ parameterized by $\theta_U$ (see line~\ref{algline:node-state-update}).
It returns the next node state of all nodes $h_{t+1} \doteq (h^v_{t+1})_{v\in V}$ and the graph observations of all agents $\psi_t \doteq (\psi^i_t)_{i\in I}$ based on the node states of all nodes $h_t$, all node observations $m_t$ and the state~$s_t$.
The only information required from $s_t$ are the graph and the mapping of agents to nodes. 
The graph observation $\psi^i_t$ of agent $i$ depends on its position in the graph and is concatenated with the agent's observation $o^i_t$ received from the environment (see line~\ref{algline:node-state-observation}).
$\hat{Q}$ and $\hat{U}$ denote that the respective gradient calculation is disabled.

During training, we sample a sequence of transitions from the replay memory and perform backpropagation through time analogous to the stored state method from recurrent experience replay~\cite{kapturowski19R2D2}.
The initial node state $h'_{j_0}$ is loaded from the replay memory and
subsequent node states in the sequence are recomputed (see lines~\ref{algline:load-node-state} and~\ref{algline:train-node-state-update}).
The Q-learning target requires graph observations for the next step, which we compute temporarily.
We then aggregate the squared error over all steps in the sequence (see line~\ref{algline:mse}) and perform gradient descent with respect to the agent's parameters $\theta_Q$ and the parameters of the message passing module $\theta_U$.

\begin{algorithm}
\caption{Independent DQN with Learned Graph Observations}\label{alg:psuedo-dqn}
\begin{algorithmic}[1]
\State Initialize replay memory $D$
\State Initialize action-value function $Q$  with weights $\theta_Q$
\State Initialize target weights $\hat{\theta}_{Q}$
\State Initialize node state update function $U$ with weights $\theta_U$
\For{episode $\leftarrow 0 \dots$}
    \State \algemph{$h_0 \leftarrow 0$} \Comment{Initialize node states}
    \State Obtain $s_0,\, o_0,\,\text{and } \algemphmath{m_0}$ by resetting the environment
    \For{$t \leftarrow 0 \text{ to } T-1$}
        \State \algemph{$h_{t+1}, \psi_t \leftarrow$ $\hat{U}(h_t, m_t, s_t; \theta_U)$} \label{algline:node-state-update} \Comment{Node state update}
        \For{$i \in I$}
            \State Select random action $a^i_t$ with probability $\epsilon$
            \State otherwise select action
            \Statex \hskip1.0em $a^i_t = \text{argmax}_a \hat{Q}(\algemphmath{o^i_t \mathbin\Vert \psi^i_t}, a; \theta_Q)$ \label{algline:node-state-observation}
        \EndFor
        \State Perform environment step with actions $a_t$ and get
        \Statex reward $r_t$, state $s_{t+1}$, obs $o_{t+1}$, node obs $\algemphmath{m_{t+1}}$
        \State Store $(\algemphmath{h_t, h_{t+1}, m_t, m_{t+1}, s_t, s_{t+1}}, o_t, a_t, r_t, o_{t+1})$ in $D$
        \State Initialize loss $L \leftarrow 0$
        \For{\algemph{batch sequence indices in $D$} j $\leftarrow j_0 \text{ to } j_0 + (J - 1)$}
            \If{$j=j_0$}
                \State $h'_j \leftarrow h_j$ \label{algline:load-node-state} \Comment{Load node state from replay memory}
            \EndIf
            \State $h'_{j+1}, \psi'_{j} \leftarrow U(h'_j, m_j, s_j;\theta_U)$ \Comment{Train node state update} \label{algline:train-node-state-update}
            \State $h''_{j+2}, \psi''_{j+1} \leftarrow \hat{U}(h'_{j+1}, m_{j+1}, s_{j+1}; \theta_U)$ \Comment{Target input}
            \State $y_j \leftarrow r_j + \mathbf{Z}_j \gamma \max_a \hat{Q}(\algemphmath{o_{j+1} \mathbin\Vert \psi''_{j+1}}, a;\hat{\theta}_Q)$
            \Statex with $\mathbf{Z}^i_j = \begin{cases}
            0 & \text{if agent } i \text{ is done at step } j+1\\
            1 & \text{otherwise} 
            \end{cases}$ 
            \State $L \leftarrow L + (y_j - Q(\algemphmath{o_j \mathbin\Vert \psi'_j}, a_j; \theta_Q))^2$
            \label{algline:mse}
        \EndFor
        \State Perform gradient descent on $L$ with respect
        \Statex to parameters $\theta_Q$ and $\algemphmath{\theta_U}$
        \State Update target weights $\hat{\theta}_Q$
    \EndFor
\EndFor
\end{algorithmic}
\end{algorithm}
\vspace{-0.25cm}

\section{Experiment Setup}
\label{sec:experiment-setup}
We evaluate our approach in diverse graphs based on a routing environment.
The following sections describe considered models, algorithms and graphs with greater detail.
Then we briefly describe the routing environment and a simplified supervised learning task.

\subsection{Models and Training Algorithms}
\label{sec:models-algorithms}
Our design consist of two parts, a model that generates graph observations and a reinforcement learning agent.

\paragraph{Graph Observations}
The core of the graph observations is the message passing framework, as described in Sec.~\ref{sec:message-passing}.
Any graph neural network can be used to generate such graph observations.
We use our proposed architecture from Sec.~\ref{sec:design-architecture} and consider three baseline graph neural network architectures from related work with implementations by PyTorch Geometric~\cite{fey19pytorchGeometric} and PyTorch Geometric Temporal~\cite{rozemberczki2021pytorchGeometricTemporal}.
Two architectures are feed-forward graph neural networks without recurrency.
GraphSAGE \cite{hamilton17GraphSage} is a GNN with multiple graph convolutional layers that use individual parameters. 
A-DGN \cite{gravina23antisymdgn} aims to improve learning long-range dependencies with an added diffusion term and performs multiple iterations with the same parameters.
As a recurrent baseline, GCRN-LSTM \cite{seo18gconvlstm} combines an LSTM with Chebyshev spectral graph convolutions~\cite{defferrard16chebconv}.
While our architecture uses a single sum to aggregate hidden states, GCRN-LSTM utilizes 8 Chebyshev convolutions to aggregate intermediate computations of an LSTM cell.

We define the readout function $\Psi$ of aggregated node state update information $H^v$ (see Sec.~\ref{sec:message-passing}) to graph observations $\psi^i$ as a concatenation of the current node state $h^v_K$ and the last node states $h^w_{K-1}$ that this node received from its neighbors.
This serves as a skip connection over the last iteration.
Note that no additional message exchange is necessary for this skip connection.
We apply the same readout function to all graph observation methods.

\paragraph{Agents}
We consider independent DQN~\cite{mnih2015DQN}, recurrent DQN (DQNR) \cite{kapturowski19R2D2}, CommNet \cite{sukhbaatar16commnet} and DGN\footnote{Not to be confused with the graph neural network A-DGN.} 
\cite{jiang20gcrl}.
We build upon the implementation of DGN\footnote{\url{https://github.com/PKU-RL/DGN/}, including the PyTorch version.} and reimplement the remaining approaches.
All variants share the same training setup but differ in the agent's architecture.
DQN is a feed-forward network with fully-connected layers that is trained with a Q-learning loss.
DQNR adds an LSTM~\cite{schmidhuber1997lstm} layer and is trained on sequences.
Both approaches do not feature any information exchange between agents, their policies are completely separated during execution.
CommNet extends DQNR with two communication rounds where agents exchange their hidden states before selecting an action.
DGN extends DQN with two communication rounds using self-attention~\cite{vaswani2017Attention} and adds a regularization term to the loss.
Within one communication round of CommNet and DGN, agents communicate with other agents that reside on the same node or on a node in their direct neighborhood.

\subsection{Graph Generation and Overview}
\label{sec:network-topologies}

\begin{figure*}[!t]
    \centering
    \hfill
    \newcommand{\topologyfigheight}{4.0cm}
    \begin{subfigure}[t]{0.69\textwidth}
    \begin{tikzpicture}
      \node[inner sep=0pt] (a) {\includegraphics[height=\topologyfigheight,trim={0 0 3.5cm 0},clip]{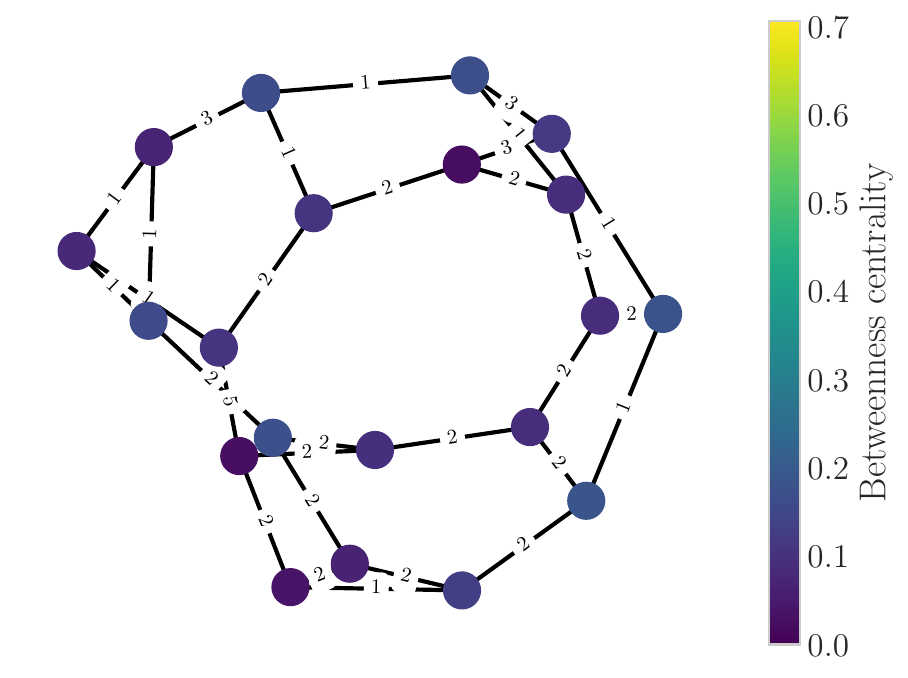}};
      \node[above right, inner sep=0pt] at ([shift={(0.8cm,0.0cm)}]a.north west) {$G_A$ \small $(0.02,\, 0.19,\ 0.11)$};
      \node[inner sep=0pt] (b) [right= -0.4cm of a] {\includegraphics[height=\topologyfigheight,trim={0 0 3.5cm 0},clip]{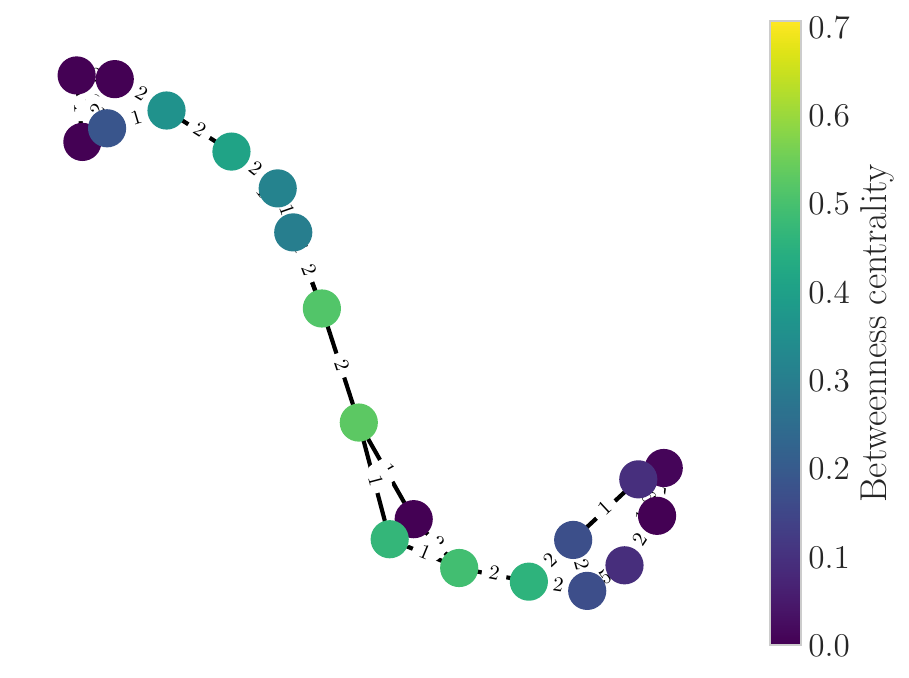}};
      \node[above right, inner sep=0pt] at ([shift={(0.8cm,0.0cm)}]b.north west) {$G_B$ \small $(0.00,\, 0.53,\, 0.23)$};
      \node[inner sep=0pt] (c) [right= -1.0cm of b] {\includegraphics[height=\topologyfigheight,trim={0 0 3.5cm 0},clip]{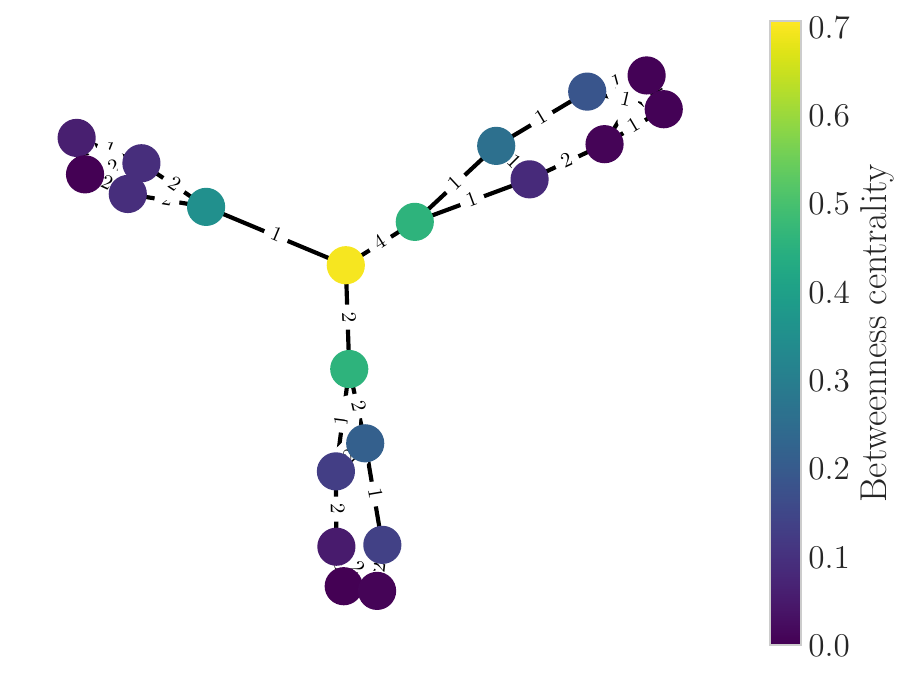}};
      \node[above right, inner sep=0pt] at ([shift={(0.8cm,0.0cm)}]c.north west) {$G_C$ \small $(0.00,\, 0.70,\, 0.16)$};
      \node[inner sep=0pt] (d) [right= -0.3cm of c] {\includegraphics[height=\topologyfigheight,trim={12cm 0 0 0},clip]{fig/graphs/betweenness_centrality_max_q1.00.pdf}};
    \end{tikzpicture}
    \caption{Exemplary test graphs $G_A$, $G_B$, and $G_C$ with increasing maximum betweenness centrality.\\ The suffix indicates the (min, max, mean) betweenness centrality in the respective graph.}
    \label{fig:example-graphs}
    \end{subfigure}
    \hfill
    \begin{subfigure}[t]{0.30\textwidth}
    \centering
    \includegraphics[height=\topologyfigheight]{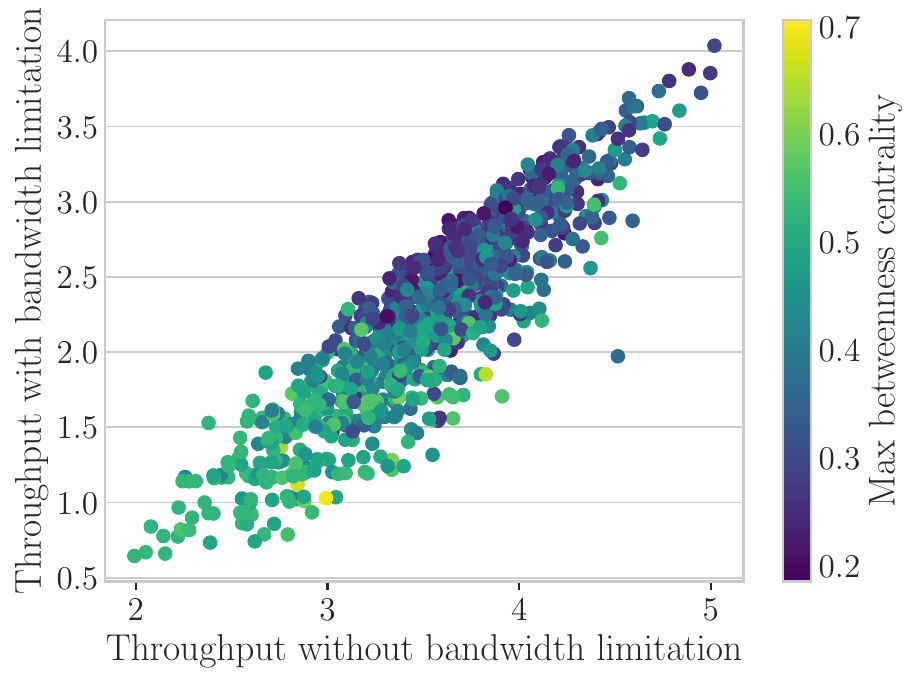}
    \caption{Mean throughput on 100 episodes for all test graphs. Each dot represents a graph.}
    \label{fig:shortest-path-routing}
    \end{subfigure}
    \hfill
    \caption{Overview of the considered graphs with (a) three exemplary graphs from the test set and (b) the mean throughput of shortest paths routing with and without bandwidth limitation in all 1000 test graphs.}
    \Description{Figure with two subfigures. Subfigure (a) shows three graphs with 20 nodes. Each node is colored according to the betweenness centrality from 0 to 0.7. The nodes on the left graph G A are dark, indicating a low betweenness centrality. The nodes of the central graph G B are lighter, indicating medium betweenness centrality. The graph on the right has a single node with high betweenness centrality in the center, the other nodes have medium betweenness centrality. Subfigure (b) shows a scatter plot of all graphs with axes for the throughput with and without bandwidth limitations. Each dot represents a graph, colored by the betweenness centrality. A positive linear relationship is visible.}
    \vspace{-0.07cm}
\end{figure*}

We extend the graph generation used in the routing environment from \citet{jiang20gcrl}.
It places $L$ nodes randomly on a 2D plane and then connects close nodes with edges until all nodes reach degree $D$.
Having a fixed node degree is not a mandatory constraint for our approach, but results in a discrete action space of fixed size that simplifies reinforcement learning.
Technically, this can be extended to graphs with nodes of variable degrees, e.g.\ via action masking~\cite{schneider2021distributed}.
The delay of an edge in steps is determined by a linear function of the distance between the connected nodes, rounded to the next integer.
Disconnected graphs are filtered out.

We generate 1000 distinct graphs for testing with $L \doteq 20$ and $D \doteq 3$.
The mean diameter is $7.21 \pm 1.42$ hops and $12.84 \pm 2.72$ steps.
The mean all-pairs shortest paths (APSP) lengths are $3.26 \pm 1.92$ hops and $5.7 \pm 3.6$ steps.
The maximum APSP lengths equal the max diameters of $12$ hops and $23$ steps.
The betweenness centrality in $[0,\, 1]$ of a node reflects the proportion of shortest paths between any two different nodes in the graph that contain this node.
In routing, high values indicate potential bottlenecks in the graph.
We show exemplary graphs with increasing maximum betweenness centrality in Fig.~\ref{fig:example-graphs}. 
The nodes are repositioned to provide a better overview.
Graph $G_A$ with a low maximum betweenness centrality is well balanced.
Graph $G_B$ has a high mean betweenness centrality due to its line-like structure.
Graph $G_C$ has a high maximum betweenness centrality because of the bottleneck node in the center.

Apart from the node connectivity and potential bottlenecks, the diameter of the graphs and the distribution of the shortest paths are expected to influence our approach.
In the graph neural network architecture considered in this paper, messages only traverse the graph through its edges.
The number of iterations for information from node $v_1$ to be forwarded to node $v_2$ equals the length of the shortest path between these nodes.
The minimum number of iterations required to collect information from all nodes is therefore the diameter of the graph.
While the maximum diameter is $12$ hops, we found that over $99\%$ of the shortest paths in all test graphs have at most $8$ hops. 
Further details are provided in the appendix.

\subsection{Routing Environment}
\label{sec:routing-environment}
\looseness=-1
We extend the routing environment from \citet{jiang20gcrl} and fix a bug that caused packets to skip edges.
At all times, there are $N$ packets of random sizes in $[0, 1)$ that have to be routed from random source to destination nodes in a given graph.
We focus on generalizability across graphs and use $N \doteq 20$ packets in our experiments.
Each packet is an agent that receives a reward of $10$ when it reaches its destination.
On a node, agents select one of $1 + D$ discrete actions that correspond to waiting and choosing an outgoing edge.
Each edge has a transmission delay given in steps.
We consider two environment modes.
The first mode has no restrictions and packets are always transmitted via their selected edges.
In the second mode, we take packet sizes and limited edge capacities into account.
A packet of size $g$ is transmitted via a selected edge if the cumulative size of all packets that are currently transmitted via this edge is smaller than $1-g$.
The packet then traverses the edge according to the number of steps in its transmission delay.
Otherwise, the packet is forced to stay at its current position and receives a penalty of $-0.2$.
An agent's local observations include its current position, its destination and packet size.
For each outgoing edge of their current node, it observes the delay, the cumulative size of packets on that edge and the respective neighbor's node id.
A node observes its own id, the number and size of packets that reside on the node, and local information about outgoing edges.
All node ids are given as one-hot encodings.

The \emph{throughput} refers to the number of packets per step that arrive at their destination. 
\emph{Delay} describes the length of their episodes.
Note that the delay should never be considered on its own.
For example, agents that only route to destinations in their $1$-hop neighborhood would achieve low delays but also a low throughput.

As a baseline, we consider heuristic agents with a global view that always choose the shortest paths with respect to the edge delays.
Fig.~\ref{fig:shortest-path-routing} shows the throughput with and without bandwidth limitations when using this heuristic for 100 episodes in all test graphs.
Each dot is colored according to the maximum betweenness centrality of the respective graph.
We can see that bandwidth limitations cause a significant drop in throughput and that graphs with high maximum betweenness tend to result in lower throughput.

\subsection{Shortest Paths Regression Task}
\label{sec:shortest-paths-regression}
The routing environment requires agents to learn paths from source to destination nodes.
To quickly evaluate the efficacy of different graph neural network architectures, we design a multi-target regression problem as a simplification of the routing environment.
We expect that the performance of different architectures in this task will indicate their suitability for the routing environment.
The training dataset contains node observations for $100\:000$ graphs generated by resetting the routing environment.
We exclude $1\:000$ of these graphs for validation.
The targets for each node are the shortest path lengths to all other nodes.
For the test dataset, we use the node observations and targets of the $1\:000$ graphs from Sec.~\ref{sec:network-topologies}.
The loss is the mean squared error between the predicted and real distances for each source and destination node.

\section{Results}
\label{sec:experiment-results}
We first present independent results for our two core components, the graph neural network architectures (see Sec.~\ref{sec:results-shortest-paths}) and agents trained in the routing environment with single graphs (see Sec.~\ref{sec:results-fixed-graph}).
Section \ref{sec:results-dynamic-graph} combines both components and provides the results for generalized routing, followed by a discussion in Sec.~\ref{sec:results-recurrent-message-passing}.

We use the AdamW optimizer \cite{loshchilov2019adamw} for all experiments. 
Details regarding the hyperparameters are provided in the appendix.

\subsection{Shortest Paths Regression}
\label{sec:results-shortest-paths}
We first evaluate the considered graph neural network architectures in the shortest paths regression task (see Sec.~\ref{sec:shortest-paths-regression}).
We train each architecture with three seeds for $50\:000$ iterations of batch size $32$.

Based on the observations regarding the APSP distribution from Sec.~\ref{sec:network-topologies}, $K=8$ message passing iterations allow to pass information between over $99\%$ of all node pairs in the test graphs.
Therefore, we hypothesize that $K = 8$ should lead to good performance on the test graphs for the non-recurrent models.
The results for different message passing iterations $K$ and unroll depths $J$ are shown in Tab.~\ref{tab:sl-results}.
In GraphSAGE, $K$ refers to the number of graph convolutional layers.
For GCRN-LSTM, we set the filter size of the Chebyshev convolutions to $K + 1$.
Both result in $K$ message passing iterations.
All approaches learn to approximate the shortest path lengths and achieve a mean squared error (MSE) of around or below $1$ for at least one configuration.
In the case of GraphSAGE, increasing the number of layers to $16$ leads to unstable training and a high test loss in this task.
As the non-recurrent architectures are stateless, they yield the same results at each forward step $t$.

For the recurrent approaches, we want to use a low number of iterations per step (i.e. $K=1$) to reduce the communication overhead.
We evaluate them with different unroll depths $J$ and higher values of $K$ for comparison.
As expected, the recurrent approaches perform poorly in the first forward step with $t=1$.
They refine their hidden states in subsequent steps and approximately reach their minimum losses at the unroll depth $J$ that was used during training.
Afterwards, we can see that their losses increase again.
Increasing the unroll depth improves long-term stability, but leads to increased training time.
A higher number of iterations $K$ predominantly leads to improved predictions, at the cost of increasing the communication overhead per step.

\begin{figure}[!t]
    \centering
    \includegraphics[width=0.9\linewidth]{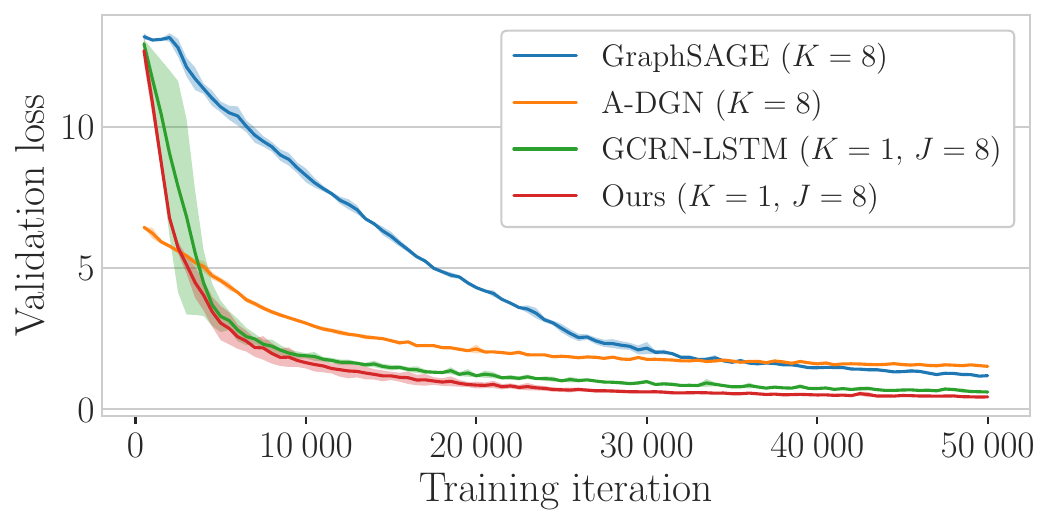}
    \caption{Validation loss of the selected GNN architectures in the shortest paths problem during training. The shaded area shows the standard deviation over 3 models.}
    \Description{The figure shows a line plot of the validation loss of selected graph neural network architectures over 50 thousand training iterations. It can be seen that the recurrent approaches with unroll depth 8 converge quickly. The non-recurrent approach A-DGN has a slightly higher loss but also converges quickly. GraphSAGE has a much higher loss and shows later convergence. However, at the last iteration, GraphSAGE has a slightly lower validation loss than A-DGN. Both recurrent approaches have a slightly lower validation loss than GraphSAGE.}
    \vspace{-0.2cm}
    \label{fig:sl-results-validation}
\end{figure}

For our experiments in combination with reinforcement learning, we select $K=8$ for the non-recurrent models and $K=1,\, J=8$ for the recurrent models.
This is a compromise between performance, stability and communication overhead.
Fig.~\ref{fig:sl-results-validation} shows the validation loss of the selected approaches during training.
The recurrent architectures converge faster to a low loss value than the non-recurrent ones.
This could possibly be caused by better gradients, as we compute separate losses for each message passing iteration when unrolling the network.
Both recurrent approaches achieve similar validation losses, although our architecture is simpler and exchanges less information during the forward steps.
The high standard deviation of GCRN-LSTM in the beginning is caused by one of the three runs, where the validation loss does not decrease initially.
In the reinforcement learning setting, we expect recurrent experience replay with stored states to further improve the long-term stability of the recurrent approaches.

\newcommand{\tablerowspacing}{2.5ex}
\newcommand{\tablevsep}{\\[0.4\medskipamount]\hline\\\\[-5.25\medskipamount]}
\begin{table}[!bt]
    \caption{Results for the shortest paths regression problem. $K$ denotes the number of message passing iterations and $J$ the unroll depth for recurrent approaches. Shown is the MSE on all test topologies after $t$ forward steps with $K$ iterations. All results are averaged over 3 seeds.}
    \small
    \centering
    \begin{tabular}{ccccccccc}
        \toprule
        \textbf{Architecture} & $K$ & $J$ & \multicolumn{6}{c}{\textbf{MSE at Forward Step $t$}}\\
        & & & 1 & 2 & 4 & 8 & 16 & 32\\
        \midrule
        \multirow{2}{*}{GraphSAGE} & 8 & - & \multicolumn{6}{c}{$1.16$ (all seeds: $1.14,\,1.22,\,1.13$)}\\
        & 16 & - & \multicolumn{6}{c}{$3.57$ (all seeds: $4.18,\,3.46,\, 3.06$)}
        \tablevsep
        \multirow{2}{*}{A-DGN} & 8 & - & \multicolumn{6}{c}{$1.50$ (all seeds: $1.49,\,1.56,\,1.46$)}\\
         & 16 & - & \multicolumn{6}{c}{$1.18$ (all seeds: $1.16,\,1.20,\,1.18$)}
        \tablevsep
        \multirow{4}{*}{GCRN-LSTM} & 1 & 8 & $4.98$ & $2.98$ & $1.12$ & $\mathbf{0.60}$ & $1.61$ & $4.27$ \\
        & 1 & 16 & $5.03$ & $3.09$ & $1.28$ & $0.60$ & $\mathbf{0.53}$ & $1.08$ \\
        & 2 & 8 & $3.18$ & $1.22$ & $0.49$ & $\mathbf{0.40}$ & $0.51$ & $1.08$\\
        & 4 & 8 & $1.56$ & $0.69$ & $0.45$ & $\mathbf{0.43}$ & $0.52$ & $1.00$
        \tablevsep
        \multirow{4}{*}{Ours} & 1 & 8 & $4.98$ & $2.91$ & $0.94$ & $\mathbf{0.43}$ & $0.99$ & $3.75$ \\
        & 1 & 16 & $5.02$ & $2.98$ & $1.02$ & $0.39$ & $\mathbf{0.37}$ & $0.46$ \\
        & 2 & 8 & $3.02$ & $1.07$ & $0.39$ & $\mathbf{0.35}$ & $0.57$ & $1.78$ \\
        & 4 & 8 & $1.26$ & $0.48$ & $\mathbf{0.34}$ & $\mathbf{0.34}$ & $0.42$ & $0.81$ \\
        \bottomrule
    \end{tabular}
    \label{tab:sl-results}
\end{table}

\subsection{Routing in Single Graphs}
\label{sec:results-fixed-graph}
Before we evaluate our method on multiple graphs, we train agents without graph observations in the routing environment using single graphs.
The agents are trained for $250\:000$ total steps with $24\:000$ iterations of batch size $32$.
Episodes are truncated after $300$ steps.
In Tab.~\ref{tab:routing-fixed-topology}, we show the results for the outlier graph from Fig.~\ref{fig:shortest-path-routing} at around $(4.5,\, 2.0)$.
The top half shows the results without bandwidth limitations, the bottom half with bandwidth limitations.
Without limitations, all methods learn the optimal shortest paths.
This is not surprising, as the graph is static and agents can locate themselves in the graph using the node id they receive in their observation.
The agents specialize on the graph.
With bandwidth limitations, the throughput drastically decreases and the delay increases.
We can see that in this graph, all learning approaches are able to outperform the shortest path solution in terms of mean reward and throughput. 
Except for DQN, the delay of arrived packets is slightly higher.
Surprisingly, the effect of communication is very small.
Using the same training setup for all agent architectures, we cannot reproduce the results from \citet{jiang20gcrl} in this particular graph and find that the performance of DQN is very close to DGN.

\begin{table}[!t]
    \caption{Results for routing in a selected graph, averaged over 1000 episodes and 3 models. The learning approaches outperform the shortest paths heuristic.}
    \small
    \centering
    \begin{tabular}{clccc}
        \toprule
        \textbf{Mode} & \textbf{Agent} & \multicolumn{3}{c}{\textbf{Metrics}} \\
        & & Reward & Delay & Throughput \\
        \midrule
        \parbox[t]{3mm}{\multirow{5}{*}{\rotatebox[origin=c]{90}{\parbox{1.6cm}{no limitation}}}}
& ShortestPath & $2.26 \pm 0.0$ & $4.39 \pm 0.0$ & $4.52 \pm 0.0$\\
        \cdashline{2-5}
        & \rule{0pt}{\tablerowspacing}DQN & $2.26 \pm 0.0$ & $4.39 \pm 0.0$ & $4.52 \pm 0.0$\\
        & \rule{0pt}{\tablerowspacing}DQNR & $2.26 \pm 0.0$ & $4.39 \pm 0.0$ & $4.52 \pm 0.0$\\
        & \rule{0pt}{\tablerowspacing}CommNet & $2.26 \pm 0.0$ & $4.39 \pm 0.0$ & $4.52 \pm 0.0$\\
        & \rule{0pt}{\tablerowspacing}DGN & $2.26 \pm 0.0$ & $4.39 \pm 0.0$ & $4.52 \pm 0.0$
        \tablevsep
        \parbox[t]{6mm}{\multirow{5}{*}{\rotatebox[origin=c]{90}{\parbox{1.5cm}{bandwidth\\ limitation}}}}
& ShortestPath & $0.88 \pm 0.0$ & $7.06 \pm 0.01$ & $1.98 \pm 0.0$\\
        \cdashline{2-5}
        & \rule{0pt}{\tablerowspacing}DQN & $1.05 \pm 0.00$ & $\textbf{6.81} \pm 0.08$ & $2.15 \pm 0.00$\\
        & \rule{0pt}{\tablerowspacing}DQNR & $1.09 \pm 0.00$ & $7.20 \pm 0.06$ & $2.22 \pm 0.01$\\
        & \rule{0pt}{\tablerowspacing}CommNet & $\mathbf{1.11} \pm 0.00$ & $7.22 \pm 0.03$ & $\mathbf{2.26} \pm 0.01$\\
        & \rule{0pt}{\tablerowspacing}DGN & $1.10 \pm 0.00$ & $7.23 \pm 0.13$ & $2.25 \pm 0.01$\\
        \bottomrule
    \end{tabular}
    \label{tab:routing-fixed-topology}
\end{table}

While we did not train agents for all 1000 test graphs, we have made similar observations for the other graphs we investigated.
Tab.~\ref{tab:routing-fixed-topology-overview} shows the throughput in the three graphs from Fig.~\ref{fig:example-graphs}, averaged over 1000 episodes and 3 models.
We omit the results without limitations, as all approaches match shortest paths with a mean throughput of around $4.05$ in graph $G_A$, $2.38$ in graph $G_B$ and $2.98$ in graph $G_C$.
For the limited bandwidth mode, we again find no notable difference between DQN and DGN in these graphs and notice that the learning approaches outperform shortest paths only for graph $G_A$.
The throughput achieved by the learning approaches is approximately on par with shortest paths for graph $G_B$ and $G_C$.

\begin{table}[!t]
    \caption{Througput for routing agents individually trained on the graphs from Fig.~\ref{fig:example-graphs} with varying betweenness centrality. Shown are the results for the bandwidth limitation mode.}
    \small
    \centering
    \begin{tabular}{lccc}
        \toprule
        \textbf{Agent} & \multicolumn{3}{c}{\textbf{Throughput}} \\
         & Graph $G_A$ & Graph $G_B$ & Graph $G_C$ \\
        \midrule
        ShortestPath & $3.20 \pm 0.00$ & $\textbf{1.53} \pm 0.00$ & $\textbf{1.02} \pm 0.00$\\
        \cdashline{1-4}
        \rule{0pt}{\tablerowspacing}DQN & $3.28 \pm 0.01$ & $1.43 \pm 0.02$ & $0.98 \pm 0.01$\\
        \rule{0pt}{\tablerowspacing}DQNR & $3.28 \pm 0.00$ & $1.47 \pm 0.01$ & $0.99 \pm 0.01$\\
        \rule{0pt}{\tablerowspacing}CommNet & $\textbf{3.31} \pm 0.00$ & $\textbf{1.53} \pm 0.00$ & $1.01 \pm 0.01$\\
        \rule{0pt}{\tablerowspacing}DGN & $3.29 \pm 0.00$ & $1.43 \pm 0.05$ & $1.00 \pm 0.00$\\
        \bottomrule
    \end{tabular}
    \label{tab:routing-fixed-topology-overview}
\end{table}

\subsection{Generalized Routing}
\label{sec:results-dynamic-graph}

This section presents the results for our learned graph observations.
We expect them to enable agents to generalize over different graphs.
As all agent architectures achieve similar results for single graphs, we select DQN as the underlying agent architecture due to its simplicity.
Instead of only receiving observations from the environment, agents now also receive graph observations from nodes they are located at.
We train graph observations and agents end-to-end with reinforcement learning on randomly generated graphs for $2.5$ million total steps, $240\:000$ iterations and batch size $32$. 
Episodes are truncated after $50$ steps to increase the number of generated graphs.
The resulting models are evaluated on our 1000 test graphs and $300$ episode steps for comparability with Sec.~\ref{sec:results-fixed-graph}.

\begin{table}[!t]
    \caption{Results for routing in 1000 test graphs for 300 steps using DQN with graph observations provided by the listed methods. An asterisk (*) indicates action masking at test time.}
    \small
    \centering
    \begin{tabular}{clccc}
        \toprule
        \textbf{Mode} & \textbf{Method} & \multicolumn{3}{c}{\textbf{Metrics}} \\
        & & Reward & Delay & Throughput \\
        \midrule
        \parbox[t]{3mm}{\multirow{6}{*}{\rotatebox[origin=c]{90}{\parbox{1.6cm}{no limitation}}}}
& ShortestPath & $\textbf{1.77} \pm 0.00$ & $5.59 \pm 0.00$ & $\textbf{3.54} \pm 0.00$ \\
        \cdashline{2-5}
        & \rule{0pt}{\tablerowspacing}GraphSAGE & $0.02 \pm 0.00$ & $\mathbf{4.49} \pm 0.35$ & $0.04 \pm 0.00$ \\
        & \rule{0pt}{\tablerowspacing}A-DGN & $1.15 \pm 0.02$ & $5.72 \pm 0.02$ & $2.29 \pm 0.04$\\
        & \rule{0pt}{\tablerowspacing}GCRN-LSTM & $1.55 \pm 0.01$ & $5.60 \pm 0.01$ & $3.09 \pm 0.01$\\
        & \rule{0pt}{\tablerowspacing}Ours & $1.56 \pm 0.03$ & $5.57 \pm 0.02$ & $3.12 \pm 0.07$\\
        & \rule{0pt}{\tablerowspacing}Ours* & $1.74 \pm 0.00$ & $5.65 \pm 0.00$ & $3.49 \pm 0.00$
        \tablevsep
        \parbox[t]{6mm}{\multirow{6}{*}{\rotatebox[origin=c]{90}{\parbox{1.5cm}{bandwidth\\ limitation}}}}
& ShortestPath & $1.05 \pm 0.00$ & $7.93 \pm 0.00$ & $2.26 \pm 0.00$\\
        \cdashline{2-5}
        & \rule{0pt}{\tablerowspacing}GraphSAGE & $0.02 \pm 0.02$ & $14.58 \pm 6.85$ & $0.08 \pm 0.05$\\
        & \rule{0pt}{\tablerowspacing}A-DGN & $0.30 \pm 0.14$ & $10.85 \pm 1.34$ &  $0.67 \pm 0.28$\\
        & \rule{0pt}{\tablerowspacing}GCRN-LSTM & $0.99 \pm 0.01$ & $8.37 \pm 0.02$ & $2.03 \pm 0.01$ \\
        & \rule{0pt}{\tablerowspacing}Ours & $1.02 \pm 0.01$ & $8.26 \pm 0.02$ & $2.10 \pm 0.02$\\
        & \rule{0pt}{\tablerowspacing}Ours* & $\mathbf{1.10} \pm 0.00$ & $\textbf{7.59} \pm 0.01$ & $\mathbf{2.38} \pm 0.01$\\
        \bottomrule
    \end{tabular}
    \label{tab:routing-random-topology}
\end{table}

\begin{figure}[!t]
    \centering
    \vspace{-0.2cm}
    \includegraphics[width=1.0\linewidth]{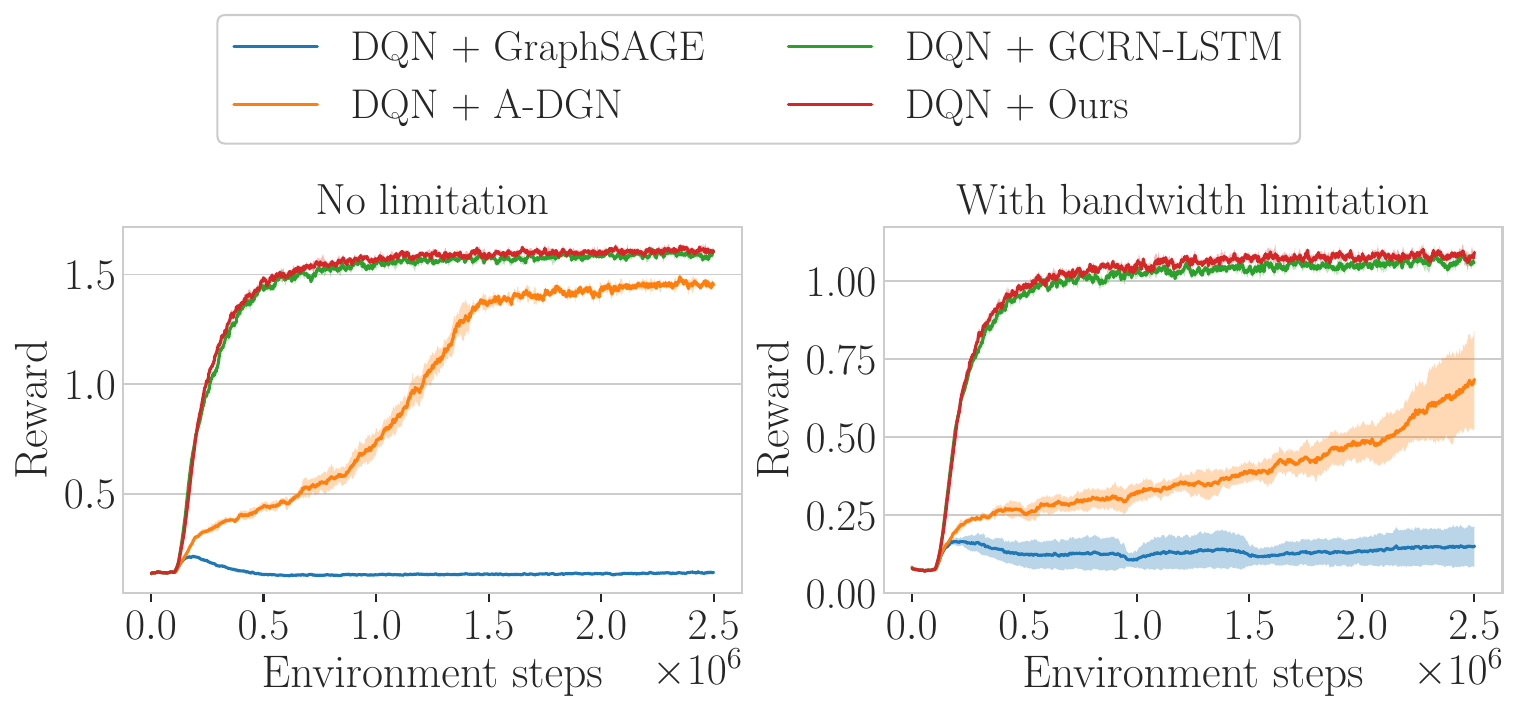}
    \vspace{-0.45cm}
    \caption{Reward of DQN with graph observations during training without (left) and with (right) bandwidth limitations. The shaded area shows the standard deviation over 3 models.}
    \Description{Figure with two line plots side by side. The left plot shows the reward over time in the environment mode without bandwidth limitations. The right plot shows the reward over time in the environment with bandwidth limitations. Both plots show the reward over 2.5 million environment steps. It can be seen that the recurrent approaches converge after around 0.75-1 million steps. Without limitations, A-DGN converges to a lower reward at around 1.5 million steps, but still increases slightly until the end. With limitations, A-DGN does not converge within 2.5 million steps but its reward increases slightly over time. GraphSAGE fails to learn in both modes, indicated by a low reward in both plots.}
    \label{fig:generalized-routing}
    \vspace{-0.25cm}
\end{figure}

Tab.~\ref{tab:routing-random-topology} shows the average results over 3 seeds. The reward during training is shown in Fig.~\ref{fig:generalized-routing}.
The high positive reward and throughput of GCRN-LSTM and our proposed architecture show that graph observations indeed enable agents to generalize over different graphs.
Our method achieves comparable results while having a lower communication overhead.
However, we find that the results are worse than for agents that specialize on single graphs (see Sec.~\ref{sec:results-fixed-graph}).
For the non-recurrent approaches, A-DGN learns graph observations but converges much slower than the recurrent approaches.
GraphSAGE fails to learn, even in experiments with batch size $256$ and jumping knowledge networks~\cite{xu2018jumpingKnowledge} that are not shown here.
Considering the results of Sec.~\ref{sec:results-shortest-paths}, it is unclear why the non-recurrent approaches perform poorly in this setting.
We hypothesize that the targets provided by backpropagation through time facilitate learning, but further experiments in different environments would be required to verify this.
All methods have a lower throughput than the shortest paths heuristic without additional modifications.

Upon closer inspection of the behavior of a model trained with our architecture, we notice that around 12\% of the 1000 test episodes contain packets that never arrive at their destination within 300 steps.
This is caused by routing loops, a common issue that can be addressed with post-processing of the learned policy \cite{hope2021GNNwithDRL}.
When repeating the experiment with different seeds, we observe routing loops in different graphs.
We investigate an action masking mechanism that stores the path of a packet and masks actions that lead to already visited nodes.
If there are no legal actions, the packet is dropped and a new packet spawns at a random location.

Action masking results in throughput improvements that match our expectations from the fixed topology setting, as shown in  Tab.~\ref{tab:routing-random-topology}.
However, it introduces $0.01$ and $0.1$ dropped packets per step for routing without and with bandwidth limitations, respectively.

\subsection{Adaptation and Limitations}
\balance
\label{sec:results-recurrent-message-passing}

In this section, we investigate how agents react to a novel situation and discuss the limitations of our work.

\begin{figure}[!t]
    \centering
    \includegraphics[width=1.0\linewidth]{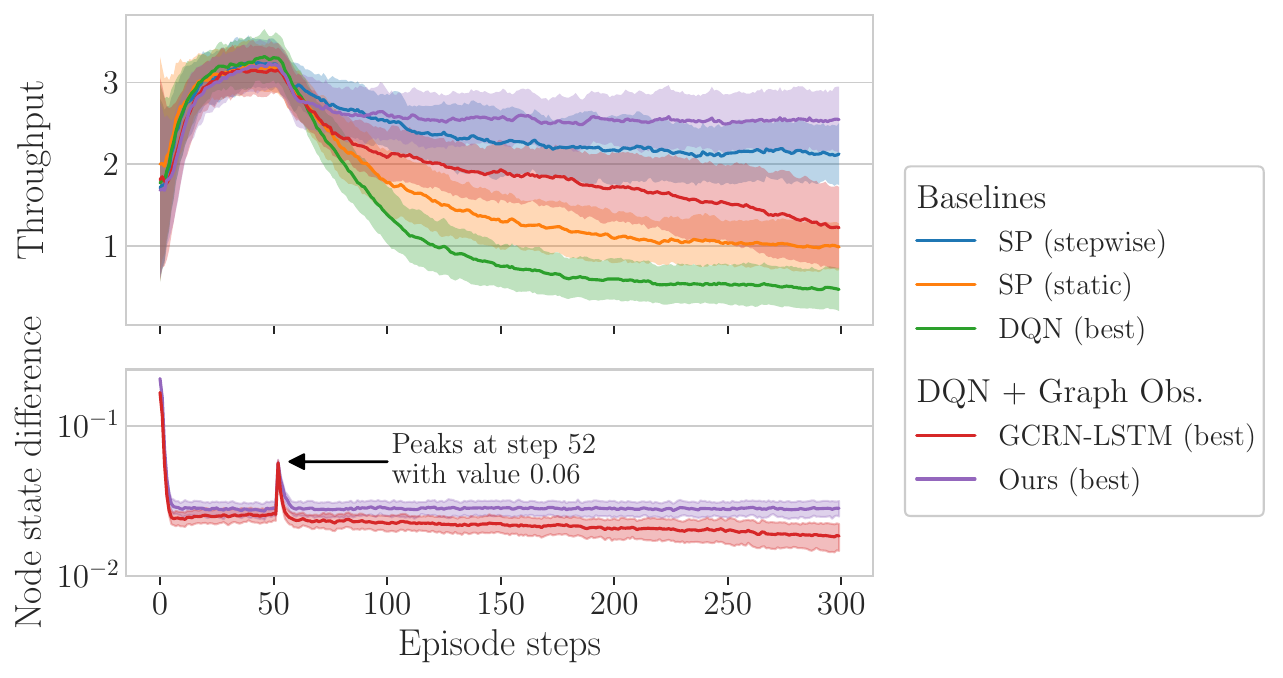}
    \vspace{-0.55cm}
    \caption{Throughput over time and node state differences of selected models in graph $G_A$ from Fig.~\ref{fig:example-graphs} averaged over 100 episodes. The delay of a single edge is increased from 2 to 10 at step 50. The shaded area shows the standard deviation.}
    \Description{The figure shows two line plots over 300 episode steps. The top one shows the throughput of the considered approaches, the bottom one the node state difference of the recurrent approaches. The throughput increases until step 50 and then drops for all approaches and the considered shortest paths baselines. DQN without graph observations shows a low throughput afterwards, lower than static shortest paths. Stepwise shortest paths has a higher higher throughput. After convergence, the combination of DQN and our approach has a higher throughput than stepwise shortest paths. The throughput of DQN and GCRN-LSTM is between static and stepwise shortest paths, but decreases rapidly within the last steps. In the bottom graph, it can be seen that the node state differences decrease quickly at the beginning of the episode and then remain almost constant. At step 52, the node state differences of both approaches peak and then quickly decrease again. In the long term, the node state differences of GCRN-LSTM decrease slowly while our architecture shows rather constant node state differences.}
    \label{fig:adaptation}
    \vspace{-0.35cm}
\end{figure}

We exemplary increase the delay of a bottleneck edge in graph $G_A$ from $2$ to $10$ at step $50$ and evaluate its effect on $100$ episodes of $300$ steps with bandwidth limitations.
Fig.~\ref{fig:adaptation} shows the throughput of different approaches in this scenario, combined with the stepwise mean absolute difference of node state values from the recurrent approaches.
Static shortest paths (SP) ignores the delay change, stepwise SP considers it.
For each learning approach, we show the results of the best model.
The recurrent approaches quickly converge to small node state differences at the beginning and react to the change at step $50$, although changing edge delays were never encountered during training.
Before step $50$, DQN performs slightly better than our approach.
After the change, all three DQN models fail to adapt and display a poor throughput, while one out of three models with our approach is able to outperform stepwise shortest paths.
Following research could consider dynamic graph changes during training and explore adaptivity in more detail.

These improvements in generalizability and adaptivity come at the cost of exchanging messages with all neighbor nodes at each step.
Fig.~\ref{fig:adaptation} shows that there is comparatively little change in the node states after convergence, suggesting a reduced need for communication.
While we show that a single message passing iteration per step suffices to learn graph observations for generalized routing, future work could investigate the further reduction of communication overhead.
We see great potential for synergies with recent works in the area of agent-to-agent communication, where agents decide when to send messages~\cite{liu20When2Com, shuai23modelBasedSparseCommunication, xuefeng23twoHopCommunication} to selected recipients instead of broadcasting them to all other agents~\cite{ma22DCC}.

\section{Related Work}
\label{sec:related-work}

Reinforcement learning for graph-based environments exemplified by routing has been investigated since the introduction of Q-learning \cite{boyan1993QLearning}.
Recent works have shown to improve over previous algorithms in various domains and network conditions \cite{nie23graphRL, li2022MARLinFutureInternet}.

Many of these approaches assume centralized control with a global view ~\cite{almasan2022gnnRouting, schneider21multiObjectiveCoordination, kim2022DRLonSDN, velasco2021RLSDN, hope2021GNNwithDRL}.
This not only limits their scalability, but also their reactivity.
Decentralized approaches \cite{schneider2021distributed, brandherm2022bigMEC} are more reactive, but their partial observability may degrade performance.
Learning directly in the target network allows agents to specialize~\cite{schneider21multiObjectiveCoordination}. However, this is challenging in practice because suboptimal actions can result in unacceptable real-world costs.
Specialized agents can also get stuck in local optima, requiring retraining from scratch if the graph or the network conditions change~\cite{bhavanasi2023gcnRouting}.

Ideally, one would pre-train agents to perform well in all graphs and network conditions and optionally fine-tune them online.
Graph neural networks have shown to enable generalization in routing scenarios  \cite{rusek2020routeNetGNNModelling, ferriolGalmes2023routeNetFermi}, as opposed to traditional models with fixed input dimensions that specialize on individual graphs \cite{kim2022DRLonSDN}.
To the best of our knowledge, related works with graph neural networks are mainly restricted to centralized approaches and agent-to-agent communication \cite{yaru21multiAgentGraphAttentionCommunication, jiang20gcrl}.
A noteworthy exception is the work by \citet{geyer2018distributedRouting}, who propose a distributed message passing scheme to learn routing in a supervised setting. 
With our work, we address the gap of generalizability over graphs in the context of multi-agent reinforcement learning with decentralized execution.

\section{Conclusion}
\label{sec:conclusion}

\looseness=-1 
In this paper, we investigate the issue of generalizability in multi-agent graph environments.
We propose to decouple learning graph representations and control by conceptually separating nodes and agents.
Nodes iteratively learn graph representations and forward local graph observations to agents, allowing them to solve tasks in the graph.
We evaluate our approach based on four graph neural network architectures across 1000 diverse graphs in a routing environment.
The results indicate that recurrent graph neural networks can be trained end-to-end with reinforcement learning and sparse rewards.
Graph observations do not only allow agents to generalize over different graphs, but also to adapt to changes in the graph without retraining.
However, having no constraints on the resulting policies can lead to deteriorated behavior compared to agents that specialize on a single graph.
This is reflected by loops in the routing environment, which we show can be alleviated with action masking.

Our contributions open up multiple avenues for future research, including the further reduction of communication overhead and the exploration of dynamically changing graphs.
The effects of graph observations in different environments are also worth investigating, especially when cooperation between agents is required.




\begin{acks}
This work has been funded by the Federal Ministry of Education and Research of Germany (BMBF) through Software Campus Grant 01IS17050 (AC3Net) and the project “Open6GHub” (grant number: 16KISK014). It has been co-funded by the German Research Foundation (DFG) in the Collaborative Research Center (CRC) 1053 MAKI.
The authors thank Amirkasra Amini for the valuable discussions.
\end{acks}



\bibliographystyle{ACM-Reference-Format} 
\bibliography{bibliography}

\clearpage
\appendix

\section*{Appendix}
\vspace{0.2cm}

\section{Architecture}
This section provides further details regarding the graph neural network and agent architectures used in this work.

\subsection{Graph Neural Networks}

Node observations are encoded using a fully connected neural network with $(d_m,\, 512,\, 256,\, d_h)$ hidden units, followed by Leaky ReLU activation functions. 
The input dimension $d_m$ is the size of the node observations. 
The output dimension $d_h$ is the hidden dimension of the respective graph neural networks. 
We set $d_h = 128$ for all experiments.
All graph neural networks use the same network to encode node observations.

\paragraph{GraphSAGE} 
We use the default configuration provided by the implementation of \texttt{GraphSAGE} in PyTorch Geometric~\cite{fey19pytorchGeometric} with $d_h = 128$ and $K$ layers. It uses the ReLU activation function between graph convolutional layers. 
Experiments with Leaky ReLU for improved consistency resulted in instabilities during training, especially for a higher number of layers.
We also experimented with jumping knowledge networks \cite{xu2018jumpingKnowledge}.
While they allowed for an improved convergence speed in the supervised setting, they did not improve learning of graph observations.

\paragraph{A-DGN} 
We use the default configuration provided by the implementation of \texttt{AntiSymmetricConv} in PyTorch Geometric~\cite{fey19pytorchGeometric} with $d_h = 128$ and $K$ iterations. 
It uses the tanh activation function between message passing iterations.

\paragraph{GCRN-LSTM} 
We use the default configuration provided by the implementation of \texttt{GConvLSTM} in PyTorch Geometric Temporal~\cite{rozemberczki2021pytorchGeometricTemporal} with $d_h = 128$ and filter size $K + 1$, resulting in $K$ message passing iterations.
The architecture uses internal hidden and cell states of size $d_h$ for each node.

\paragraph{Ours} 
We use two LSTM cells that share single hidden and cell states of size $d_h = 128$ for each node.

\subsection{Agent Architectures}
All agent architectures use a fully connected neural network with $(d_o,\, 512,\, 256)$ hidden units, followed by Leaky ReLU activation functions to encode the agent's observation.
The input dimension $d_o$ is determined by the size of the agent's observations.
Note that graph observations lead to an expanded observation space.

\paragraph{DQN}
DQN uses a single linear layer of input and output sizes $(256,\, D + 1)$ to map the observation encoding to the discrete action space with $D + 1$ actions.
For our experiments, we set $D = 3$.

\paragraph{DQNR}
The output of the observation encoder is followed by an LSTM cell with hidden and cell state of size $256$.
The action is predicted with a fully connected linear layer with input and output size $(256,\, D + 1)$, using the hidden state as input.

\paragraph{CommNet}
The architecture of CommNet builds upon DQNR and adds two rounds of message exchange using the updated hidden state of the LSTM.
One round of message exchange is the sum of the agent's hidden state and the mean of all neighbors' hidden states, excluding the agent's own hidden state.

\paragraph{DGN}
Agents perform two rounds of message exchange using self-attention.
We use hyperparameters based on the official implementation of DGN, i.e. $2$ attention layers with $8$ attention heads, and key and value size $16$.
However, we use in- and output size $256$ instead of $128$ for consistency.
The outputs of the attention layers are concatenated with the observation encoding and projected to actions using a linear layer of sizes $(3 \cdot 256,\, D + 1)$.

\section{Experiment Details}
Tab.~\ref{tab:training-testing-parameters} provides an overview of the parameters used during training and testing.
The learning rate is set to $0.001$ in the shortest paths regression task (see Sec.~\ref{sec:results-shortest-paths}).
For our evaluations, we use the models from the last training iteration.
The following sections provide further details regarding our implementation of Alg.~\ref{alg:psuedo-dqn}.

\subsection{Deep Q-Learning}
At the beginning of Alg.~\ref{alg:psuedo-dqn}, we initialize the parameters of the target action-value network as $\hat{\theta}_Q \leftarrow \theta_Q$.
As in the implementation of DGN,\footnote{\url{https://github.com/PKU-RL/DGN/}, including the PyTorch version.} the target parameters $\hat{\theta}_Q$ are then smoothly updated with $\hat{\theta}_Q' \leftarrow \tau \theta_Q + (1 - \tau) \hat{\theta}_Q$ in each iteration.
DGN further augments the regular DQN loss $(y_j - Q(o_j, a_j; \theta_Q))^2$ with a regularization term $(\hat{Q}(o_j, a; \theta_Q) - Q(o_j, a; \theta_Q))^2$ for all other actions $a \neq a_j$.

\subsection{Graph Observations}
During execution, we implement graph observations as an environment wrapper.
During training, as described in Sec.~\ref{sec:integration-dqn}, we sample a mini batch of a sequence of steps to perform backpropagation through time.
In our implementation, these sequences are allowed to cross episode boundaries.
When reaching the end of an episode, we reset the node states $h_j'$ to zero.
This is done independently for each sequence in the mini batch.

\begin{table}[h]
    \centering
    \caption{Parameters used for training and testing.}
    \begin{tabular}{lcc}
        \toprule
        \textbf{Parameter} & \multicolumn{2}{c}{\textbf{Environment Mode}} \\
          & Single Graph & Generalized \\
        \midrule
        Optimizer & \multicolumn{2}{c}{AdamW~\cite{loshchilov2019adamw}} \\
        Learning rate & \multicolumn{2}{c}{$0.001$} \\
        Total Steps & $250\:000$ & $2\:500\:000$ \\
        Steps before training & $10\:000$ & $100\:000$ \\
        Replay memory size & \multicolumn{2}{c}{$200\:000$ steps} \\
        Steps between iterations & \multicolumn{2}{c}{$10$} \\
        Episode length & $300$ steps & $50$ steps \\
        Mini batch size & \multicolumn{2}{c}{$32$} \\
        Target network update $\tau$ & \multicolumn{2}{c}{$0.01$} \\
        Discount factor $\gamma$ & \multicolumn{2}{c}{$0.9$} \\
        Initial exploration $\epsilon$ & \multicolumn{2}{c}{$1.0$} \\
        $\epsilon$-decay (per 100 steps) & $0.996$ & $0.999$ \\
        Minimum exploration $\epsilon$ & \multicolumn{2}{c}{$0.01$}
        \tablevsep
        Test exploration $\epsilon$ & \multicolumn{2}{c}{$0.0$} \\
        Test episodes & \multicolumn{2}{c}{$1\:000$}\\
        Test episode length & \multicolumn{2}{c}{$300$ steps} \\
        \bottomrule
    \end{tabular}
    \label{tab:training-testing-parameters}
\end{table}

\section{Test Graphs}
\nobalance
While the training graphs are randomly generated on the fly and therefore depend on the used training seed, the test graphs are fixed and independent of the training seed.
This section supplements Sec.~\ref{sec:network-topologies}  and provides further details regarding the test graphs. 

\subsection{Metrics}
Tab.~\ref{tab:graph-stats} lists the metrics we used to compare the graphs, including details regarding the betweenness centrality briefly mentioned in the paper. 
Considering the graph max betweenness centrality, the standard deviation of $0.1$ and the high difference between the minimum value of $0.19$ and the maximum value of $0.7$ suggest that the test set contains diverse graphs with and without bottleneck nodes.
We think that it is important for future work in graph-based environments to provide similar metrics for their test graphs to improve comparability.
One could also draw inspiration from communication networks and classify the graphs according to their structure. 
When considering dynamic graphs in future work, it will be essential to further quantify the dynamicity of the graphs.

\subsection{Shortest Paths Distribution}
\begin{figure}[b]
    \centering
    \includegraphics[width=0.9\linewidth]{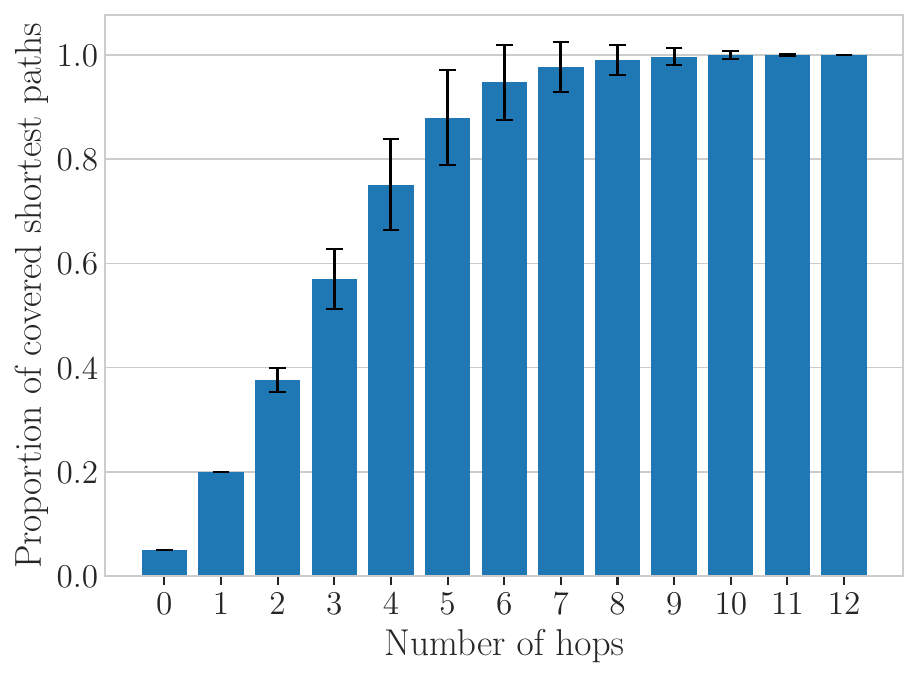}
    \caption{Cumulative distribution of APSP lengths on the test graphs. The error bars indicate the standard deviation.}
    \Description{The figure shows bars that indicate the proportion of covered shortest paths based on the number of hops. The size of the bars increases quickly, reaching over 99\% coverage after 8 hops.}
    \label{fig:topology-stats-asap}
\end{figure}

Fig.~\ref{fig:topology-stats-asap} shows the cumulative distribution of the all-pairs shortest paths (APSP) lengths of the test graphs in hops. While the longest shortest path takes 12 hops, over $99\%$ of all shortest paths take at most 8 hops.
We therefore considered $8$ to be a suitable candidate for the number of message passing iterations $K$ of the non-recurrent approaches and unroll depth $J$ of the recurrent approaches.
A higher number may lead to improved performance in theory, at the cost of a higher computational overhead during training and a higher communication overhead during execution.
Our experiments show that the effect of using more message passing iterations depends on the used architecture and is not necessarily positive (see Sec.~\ref{sec:results-shortest-paths}).

\begin{table}[b]
    \small
    \centering
    \caption{Metrics for the test graphs. Shown are the minimum, maximum, mean, and standard deviation of each metric. The first three metrics have the same values for all graphs.}
    \begin{tabular}{lcccc}
        \textbf{Metric} & \textbf{Min} & \textbf{Max} & \textbf{Mean} & \textbf{Std}\\
        \toprule
        Order (\# nodes) & \multicolumn{3}{c}{$20$} & $0$\\
        Node degree (\# incident edges) & \multicolumn{3}{c}{$3$} & $0$\\
        Size (total \# edges) & \multicolumn{3}{c}{$30$} & $0$\\
        Diameter (hops) & $5.00$ & $12.00$ & $7.21$ & $1.42$\\
        Diameter (delays) & $8.00$ & $23.00$ & $12.84$ & $2.72$\\
        APSP (hops) & $0.00$ & $12.00$ & $3.26$ & $1.92$\\
        APSP (delays) & $0.00$ & $23.00$ & $5.70$ & $3.60$\\
        Node betweenness centrality & $0.00$ & $0.70$ & $0.15$ & $0.12$\\
        Graph max betweenness centrality & $0.19$ & $0.70$ & $0.39$ & $0.10$\\
        Graph mean betweenness centrality & $0.11$ & $0.23$ & $0.15$ & $0.02$\\
        \bottomrule
    \end{tabular}
    \label{tab:graph-stats}
\end{table}

\end{document}